\documentclass[fleqn,useAMS,usenatbib]{mnras}
\usepackage{amsmath}
\usepackage{aecompl}
\usepackage{txfonts}
\usepackage{natbib}
\usepackage{aas_macros}
\usepackage[all]{xy}
\usepackage{booktabs}
\usepackage{multirow}

 \usepackage{ifthen}
 \ifx\pdftexversion\undefined
 \usepackage[dvips]{graphicx}
 \else
 \usepackage[pdftex]{graphicx}
 \usepackage{epstopdf}
 \usepackage{thumbpdf}
 \fi

\usepackage[all]{hypcap}

\def\del#1{{}}

\newcommand{\dd}{\mathrm{d}}
\newcommand{\ellvec}{\bmath{\ell}}

\title[Bias to CMB lensing-galaxy shape cross-correlation]{Imitating intrinsic alignments: A bias to the CMB lensing-galaxy shape cross-correlation power spectrum induced by the large-scale structure bispectrum}
\author[Philipp M. Merkel and Bj{\"o}rn Malte Sch\"afer]
{Philipp M. Merkel$^1$\thanks{e-mail: philipp.merkel@urz.uni-heidelberg.de} and Bj{\"o}rn Malte Sch\"afer$^2$\\
${}^1$Institut f{\"u}r Theoretische Astrophysik, Zentrum f{\"u}r Astronomie, Universit{\"a}t Heidelberg, Philosophenweg 12, 69120 Heidelberg, Germany\\
${}^2$Astronomisches Recheninstitut, Zentrum f{\"u}r Astronomie, Universit{\"a}t Heidelberg, Philosophenweg 12, 69120 Heidelberg, Germany}

\begin{document}
\pagerange{\pageref{firstpage}--\pageref{lastpage}}
\pubyear{2017}
\maketitle
\label{firstpage}

\begin{abstract}
Cross-correlating the lensing signals of galaxies and comic microwave background (CMB) fluctuations is expected to provide valuable cosmological information. In particular it may help tighten constraints on parameters describing the properties of intrinsically aligned galaxies at high redshift. To access the information conveyed by the cross-correlation signal its accurate theoretical description is required. We compute the bias to CMB lensing-galaxy shape cross-correlation measurements induced by nonlinear structure growth. Using tree-level perturbation theory for the large-scale structure bispectrum we find that the bias is negative on most  angular scales, therefore mimicking the signal of intrinsic alignments. Combining \textit{Euclid}-like galaxy lensing data with a CMB experiment comparable to the \textit{Planck} satellite mission the bias becomes significant only on smallest scales ($\ell\gtrsim 2500$). For improved CMB observations, however, the  corrections amount to 10-15~per~cent of the CMB lensing-intrinsic alignment signal over a wide multipole range~($10 \lesssim \ell \lesssim 2000$). Accordingly the power spectrum bias, if uncorrected, translates into~$2\sigma$ and~$3\sigma$ errors in the determination of the intrinsic alignment amplitude in case of CMB stage~III and stage~IV experiments, respectively.
\end{abstract}


\section{Introduction}
\label{sec_introduction}
Two fundamental observational pillars of the nowadays well-established cosmological standard model are the cosmic microwave background (CMB) and gravitational lensing. Fluctuations in CMB temperature and polarization have preserved the physical conditions in the early Universe while gravitationally deflected light of distant galaxies unveils the Universe's late-time evolution. This complementarity provides a powerful tool to constrain cosmological parameters \citep{2002PhRvD..65b3003H}.

Gravitational light deflection by the intervening large-scale structure, however, is by no means limited to galaxies but applies to CMB radiation, too. Since CMB photons emanate from the last-scattering surface, which resides at much higher redshift than galaxies do, both photons of galaxies and the CMB traverse the same lensing matter structures before reaching the contemporary observer. Consequently, the signals of CMB lensing and cosmic shear are correlated. This cross-correlation has been successfully detected in various data sets \citep{2015PhRvD..91f2001H,2015PhRvD..92f3517L,2016MNRAS.460..434H,2016MNRAS.459...21K,2017MNRAS.464.2120S}. In addition to that it is expected that CMB lensing is correlated with intrinsic galaxy shapes \citep{2014MNRAS.443L.119H,2014PhRvD..89f3528T} because the intrinsic ellipticity of a galaxy is presumably determined through tidal interactions exerted by the surrounding large-scale structures \citep{2001MNRAS.320L...7C}, which in turn contribute to the lensing signal of the CMB. Intrinsically aligned galaxies are a major contaminant to cosmic shear measurements and have been observed in several weak lensing surveys \citep[see][for an overview and references therein]{2015SSRv..193..139K}. Conversely, intrinsic alignments can be used to probe tidal fields in the large-scale structure \citep{2013JCAP...12..029C}. In both cases a thorough understanding of intrinsic alignments is required.

While cosmic shear becomes manifest in an elongation or compression of the apparent galaxy shape the weak lensing effect on the CMB is more subtle and not directly accessible to observations. Lensing, however, changes the statistical properties of the CMB in a characteristic way: the lensed CMB  is statistically anisotropic, which can be used to reconstruct the lensing signal from observed temperature and polarization maps by means of quadratic estimators under the assumption of statistical homogeneity of the unlensed CMB \citep{2001PhRvD..64h3005H,2002ApJ...574..566H,2003PhRvD..67h3002O}. By construction the estimators are unbiased, i.e. their expected values provide a faithful reconstruction of the lensing signal. But this does not necessarily hold true for higher statistics and several biases to its estimated power spectrum have been identified \citep{2003PhRvD..67l3507K,2011PhRvD..83d3005H,2013PhRvD..88h3517A,2013MNRAS.429..444M,2016PhRvD..94d3519B}. In this work we investigate biases in measurements of the CMB lensing-cosmic shear and CMB lensing-intrinsic ellipticity cross-correlations, respectively. We quantify their strength and examine how they vary with CMB data quality. The respective power spectrum biases we are considering arise from the large-scale structure bispectrum, which is non-zero because nonlinear clustering makes the distribution of matter fluctuations asymmetric. We restrict our analysis to the temperature based estimator of the CMB lensing effect and defer its extension to polarization to future work.

Identification and quantification of potential biases is particularly important as the additional information contained in the CMB lensing-galaxy shape cross-correlations may be exploited to further improve cosmological parameter constraints \citet{2015MNRAS.449.2205K}. Especially its potential to shed light on intrinsic galaxy alignments at high redshift is of rather great interest \citep{2014MNRAS.443L.119H,2014PhRvD..89f3528T}. Furthermore detections of the CMB lensing-cosmic shear cross-correlation report consistently an amplitude which is lower than predicted by the fiducial cosmological model. The discrepancy is statistically insignificant and to some extent it can be explained by contributions from intrinsic alignments \citep{2014MNRAS.443L.119H,2014PhRvD..89f3528T,2015MNRAS.453..682C}, tough it is worthwhile to further elucidate its possible origin.  

This article has the following structure: In Section~\ref{sec:formalism} we derive expressions for the bias to CMB lensing-galaxy shape cross-correlation power spectra. We present results for a variety of different CMB experiments in Section~\ref{sec:results} and conclude in Section~\ref{sec_conclusion}. The appendix gathers details on the large-scale structure bispectrum computation employed in this work.

All our results are illustrated using a spatially flat $\Lambda\mathrm{CDM}$ cosmology the specific parameter values of which are compatible with the latest data release of the \citet{2016A&A...594A..13P}: $\Omega_\mathrm{m}=0.312$, $\Omega_\mathrm{b}=0.0483$, $\sigma_8=0.834$, $n_\mathrm{s}=0.9619$ and $h=0.67556$ for the matter and baryon density, respectively, the normalization and spectral index of the power spectrum, and the Hubble parameter evaluated today. For the computation of dark matter and CMB power spectra we make use of the \textsc{cosmic linear anisotropy solving system} \citep[\textsc{class};][]{2011JCAP...07..034B}; the numerical evaluation of the bias expressions has been realized using the integration methods provided by the \textsc{cuba} library \citep[][]{2005CoPhC.168...78H}.

\section{Formalism}
\label{sec:formalism}

\subsection{Extrinsic and intrinsic galaxy ellipticities}
\label{subsec:extrinsic_and_intrinsic_ellipticities}

The shape of a distant galaxy is characterized by its (complex) ellipticity. As long as lensing effects are small the observed ellipticity is the sum of intrinsic galaxy ellipticity~$\epsilon$ and cosmic shear~$\gamma$
\begin{equation}
 \epsilon^\mathrm{obs} \simeq \epsilon + \gamma = \epsilon_+ + \gamma_+ + \mathrm{i} \left( \epsilon_\times + \gamma_\times  \right)
\end{equation}
(see \citealp{2001PhR...340..291B} and \citealp{2015PhR...558....1T,2015SSRv..193...67K} for reviews on gravitational lensing and intrinsic galaxy ellipticities, respectively).
We use the so-called linear model to describe intrinsic ellipticities \citep{2001MNRAS.320L...7C,2004PhRvD..70f3526H}. The linear model is supposed to primarily apply to elliptical galaxies while for spirals more complex models exist \citep{2001ApJ...559..552C, 2002MNRAS.332..788M}. In its framework galaxies are shaped by the tidal shear field of the ambient dark matter structures well within the era of matter domination.
Both intrinsic and lensing induced ellipticity can be derived as second derivatives of the Newtonian gauge potential~$\Phi$ modulated by an appropriate weight function
\begin{equation}
 \epsilon (\bmath{\hat{n}}), \gamma (\bmath{\hat{n}}) 
 	= \left[ \left(\upartial^2_x - \upartial^2_y \right) + 2 \mathrm{i} \upartial_x\upartial_y \right] \int_0^\infty \dd\chi \,
			W_{\epsilon,\gamma} (\chi) \, \Phi (\bmath{\hat{n}}, \chi).
\end{equation}
Here we made the line-of-sight pointing in $z$-direction and assumed a flat sky which is a reasonable simplification for cosmic shear studies \citep{2008PhR...462...67M,2010CQGra..27w3001B}. Positions on the sky are indicated by the 2D vector~$\bmath{\hat{n}}$, while~$\chi$ denotes the comoving distance. The weight functions have to take the distribution of source galaxies~$n(\chi)$ accessible by the survey into account:
\begin{gather}
W_\epsilon(\chi) = - \mathcal{A}_I \,n(\chi),\\
 W_\gamma (\chi)= \chi \int_\chi^\infty \dd\chi' \frac{\chi-\chi'}{\chi'} n(\chi').
 \label{eq:cosmic_shear_weight_function}
\end{gather}
The amplitude of the intrinsic ellipticities~$\mathcal{A}_I \simeq 8.93 \times 10^{-3} H_0^{-2}$ \citep{2004PhRvD..70f3526H, 2007NJPh....9..444B, 2011A&A...527A..26J,2012MNRAS.424.1647K,2013MNRAS.434.1808M,2015MNRAS.449.2205K} is set in such a way that it matches low-$z$ SuperCOSMOS observations \citep{2002MNRAS.333..501B}.
To be specific we choose a \textit{Euclid}-like \citep{2011arXiv1110.3193L} cosmic shear experiment as reference throughout this work. It is parametrized in redshift by
\begin{equation}
 n(z)\, \dd z \sim z^2 \, \exp \left[ - \left( \frac{z}{0.64}\right)^{1.5}\right] \, \dd z 
\end{equation}
with a galaxy density of 30 galaxies per $\mathrm{arcmin}^{2}$ \citep{2013LRR....16....6A}.

\subsection{CMB lensing}
\label{subsec:CMB_lensing}

CMB temperature fluctuations are characterized by the temperature contrast
\begin{equation}
	\Theta (\bmath{\hat n}) \equiv \frac{T (\bmath{\hat n}) - \bar T}{\bar T},
\end{equation}
i.e. by the relative deviation from the mean CMB temperature.
The effect of lensing on the CMB is a remapping of these temperature fluctuations \citep[see][for a review]{2006PhR...429....1L}
\begin{equation}
 \tilde\Theta ( \bmath{\hat{n}} ) = \Theta ( \bmath{\hat{n}} + \nabla \phi ) \simeq \Theta ( \bmath{\hat{n}} ) 
 				+ \nabla^a \Theta \nabla_a\phi 
				+ \frac{1}{2} \nabla_a\nabla_b\Theta\nabla^a \phi \nabla^b\phi
				+\hdots
\label{eq:lensed_CMB_temperature}
\end{equation}
according to the gradient of the lensing potential
\begin{equation}
 \phi(\bmath{\hat{n}}) =  \int_0^{\chi^\star}\dd\chi \, W_\mathrm{CMB} (\chi) \, \Phi(\bmath{\hat{n}}, \chi) 
\end{equation}
where
\begin{equation}
 W_\mathrm{CMB} (\chi) = 2\,\frac{\chi^\star - \chi}{\chi\chi^\star}.
 \label{eq:CMB_lensing_weight_function}
\end{equation}
Repeated indices are summed over and we denote the comoving distance to the infinitely thin surface of last scattering by~$\chi^\star$.

\begin{figure}
 \resizebox{\hsize}{!}{\includegraphics{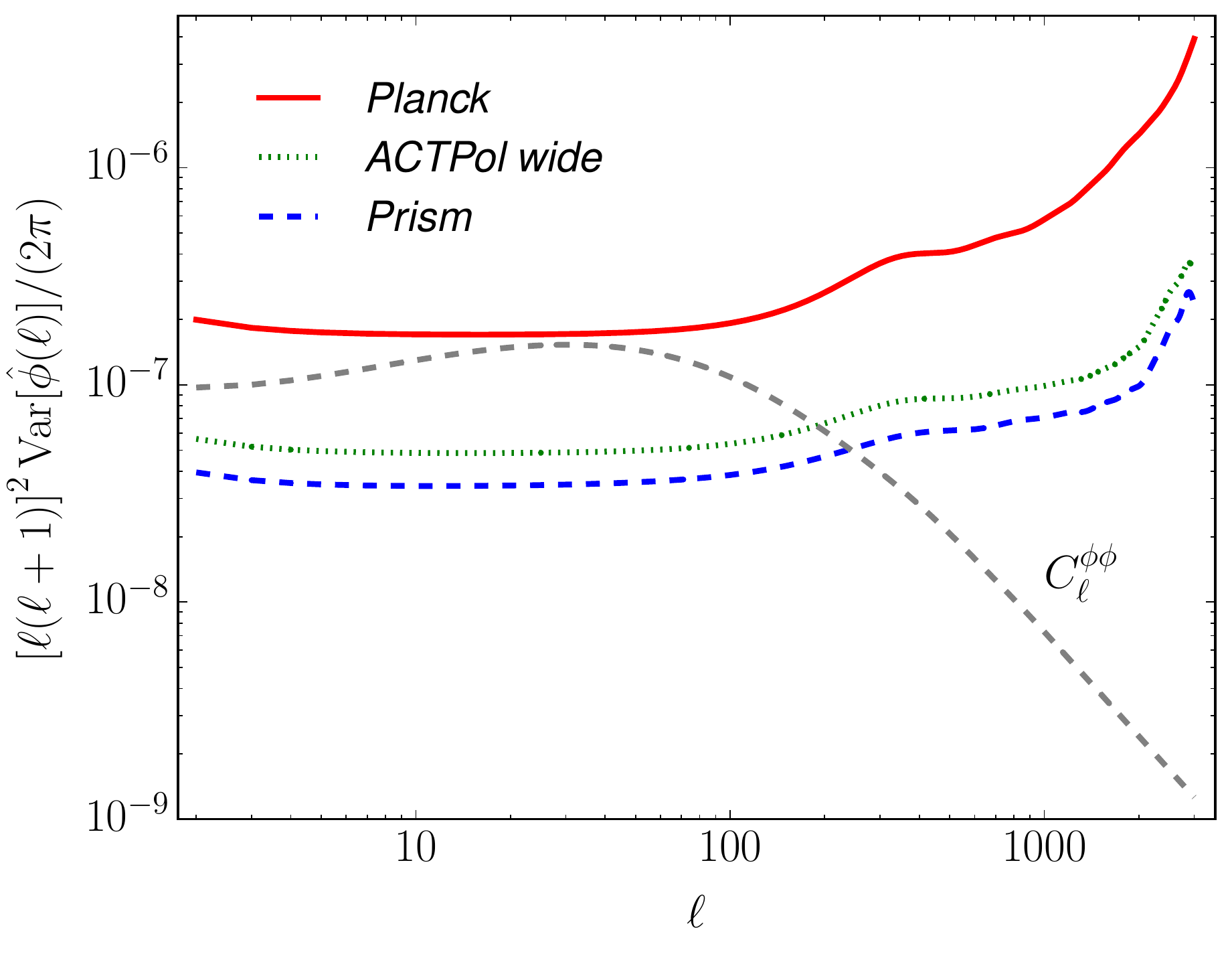}}
 \vspace{-15pt}
 \caption{Variance of the CMB lensing potential estimator based on CMB temperature observations for the three different experiments considered in this work. The CMB lensing power spectrum (grey dashed line) is plotted as reference.}
 \label{fig:estimator_variance}
\end{figure}

The lensing potential itself is not observable but it may be statistically reconstructed from the observed, i.e. lensed temperature fluctuations by means of a quadratic estimator \citep{2001PhRvD..64h3005H}. Its harmonic space representation in the flat-sky limit reads
\begin{equation}
 \hat\phi (\bmath \ell ) = A_\ell \, \int\frac{\dd^2 \ell_1}{(2\upi)^2} \, g ( \bmath\ell_1, \bmath \ell) \tilde\Theta (\bmath\ell_1) \tilde\Theta ( \bmath \ell - \bmath\ell_1).
 \label{eq:CMB_lensing_potential_estimator}
\end{equation}
The weight function
\begin{equation}
 g(\bmath\ell_1, \bmath \ell_2 ) = \frac{f (\ell_1, \ell_2)}{2 C^{\Theta\Theta}_{\ell_1,\mathrm{obs}} C^{\Theta\Theta}_{\ell_2,\mathrm{obs}}}
 						= \frac{( \bmath\ell_1 + \bmath\ell_2 ) \cdot  \left( \bmath\ell_1  C^{\Theta\Theta}_{\ell_1}
							+ \bmath\ell_2 C^{\Theta\Theta}_{\ell_2}\right) }{2 C^{\Theta\Theta}_{\ell_1,\mathrm{obs}} C^{\Theta\Theta}_{\ell_2,\mathrm{obs}}}
\label{eq:estimator_weight_function}
\end{equation}
is chosen such that the (Gaussian) variance of the estimator is minimal, while the normalization
\begin{equation}
A^{-1}_\ell = \int\frac{\dd^2 \ell_1}{(2\upi)^2} f(\ellvec_1, \ellvec-\ellvec_1) \, g ( \ellvec_1, \ellvec - \ellvec_1 )
\label{eq:estimator_normalization}
\end{equation}
ensures that the estimator is unbiased. The observed CMB temperature power spectrum contains both the lensing signal and instrumental noise. The latter is characterized by a Gaussian beam width~$\sigma_\mathrm{FWHM}$ and the standard deviation of white Gaussian pixel noise~$\sigma_\mathrm{N}$ \citep{2002PhRvL..89a1303K}
\begin{equation}
  C^{\Theta\Theta}_{\ell,\mathrm{obs}} = C^{\tilde\Theta\tilde\Theta}_\ell + \left(\frac{\sigma_\mathrm{N}}{T_\mathrm{CMB}}\right)^2 
  \exp\left[ \frac{\ell(\ell + 1 ) \sigma_\mathrm{FWHM}^2}{8\log 2}  \right].
  \label{eq:CMB_noise}
\end{equation}
We detail the specification of three different ongoing and forthcoming CMB experiments in Table~\ref{tab:CMBExperiments}.
\begin{table}
\caption{Specification of CMB experiments.}
\label{tab:CMBExperiments}
	\begin{tabular}{lccc}
\toprule
& $\nu$ & $\sigma_\mathrm{FWHM}$ & $\sigma_\mathrm{N}$\\
& $ [\mathrm{GHz}]$ & $[\mathrm{arcmin}]$ & $[\mu\mathrm{K\, arcmin}]$ \\
\midrule
\multirow{2}{*}{\textit{Planck}} & 143 & 7.1 & 42.60\\
	   & 217 & 5.0 & 65.50\\
\midrule
\textit{ACTPol wide} & 150 & 1.4 & 20.00\\
\midrule
\multirow{7}{*}{\textit{Prism}} & 90 & 5.7 & 18.80\\
		    & 105 & 4.8 & 13.80 \\
		    & 135 & 3.8 & 9.85 \\
		    & 160 & 3.2 & 7.78 \\
		    &185  & 2.8 & 7.05 \\
		    & 200 & 2.5 & 6.48 \\
		    & 220 & 2.3 & 6.26 \\
\bottomrule
	\end{tabular}

\medskip
Where several frequency bands are available the inverse weighted sum of all bands is taken in equation~\eqref{eq:CMB_noise}.
\end{table}
To illustrate how the CMB data quality improves from \textit{Planck} \citep{2006astro.ph..4069T} over the \textit{ACTPol wide} experiment \citep{2014PTEP.2014fB110N} to the future \textit{Prism} mission \citep{2014JCAP...02..006A} we plot the variance of the estimator~\eqref{eq:CMB_lensing_potential_estimator} in Figure~\ref{fig:estimator_variance} for all three experiments. The last two correspond to stage~III and stage~IV CMB experiments \citep{2006astro.ph..9591A}. Note that the estimator variance is identical to its normalization~\eqref{eq:estimator_normalization}.

\subsection{CMB lensing galaxy shape cross-correlation power spectra}
\label{subsec:CMB_lensing_galaxy_shape_cross_correlation}

Since both gravitational lensing of galaxies and of CMB temperature fluctuations are due to the very same large-scale structures their cross-correlation is non-zero and can be related to the three-dimensional power spectrum of dark matter fluctuations. To assess the statistical properties of the shear field components we use their parity conserving $E$-mode \citep{1998MNRAS.301.1064K}, which can be shown to be identical to the weak lensing convergence \citep{2010CQGra..27w3001B}. The two-dimensional Limber-projected \citep{1953ApJ...117..134L} power spectrum of the CMB lensing-cosmic shear cross-correlation is then given by
\begin{equation}
 C_\ell^{\gamma\phi} = \int_0^\infty \dd\chi \, \ell^{-2} \, W_\gamma(\chi) \, W_\mathrm{CMB}(\chi) 
 					\, \eta^2(\chi) \, P^\mathrm{nl}_{\delta\delta} (\ell/\chi,\chi)
 \label{eq:CMB_lensing_cosmic_shear_power_spectrum}
\end{equation}
\citep[cf.][who use the reconstructed CMB lensing convergence instead of the CMB lensing potential, thus our expression contains an additional factor $\propto \ell^{-2}$]{2015PhRvD..91f2001H}. We include small-scale corrections to the matter spectrum~$P^\mathrm{nl}_{\delta\delta}$ due to nonlinear clustering by using an extended halofit approach \citep{2003MNRAS.341.1311S,2012ApJ...761..152T} and give the explicit expression of the Poisson factor~$\eta$ in equation~\eqref{eq:Poisson_equation}. The overlap of the weight functions~$W_\gamma$ and~$W_\mathrm{CMB}$ is rather small, resulting in a comparatively weak cross-correlation between CMB lensing and cosmic shear (cf. Figure~\ref{fig:CMB_lensing_cosmic_shear_bias}).

Similarly, there is a non-vanishing cross-correlation between CMB lensing and intrinsic ellipticities:
\begin{equation}
 C_\ell^{\epsilon\phi} = \int_0^\infty\dd\chi \, \ell^{-2} \, W_\epsilon(\chi) \, W_\mathrm{CMB}(\chi) \, \eta^2(\chi) \, 
 	\frac{\sqrt{P^\mathrm{nl}_{\delta\delta} (\ell/\chi,\chi)\, P^{\mathrm{lin}}_{\delta\delta} (\ell/\chi,\chi)}}{\bar D_+(\chi)}.
 \label{eq:CMB_lensing_itrinsic_alignment_power_spectrum}
\end{equation}
(\citealp{2014MNRAS.443L.119H,2014PhRvD..89f3528T}; see \citealp{2016MNRAS.461.4343L} for quadratic ellipticity models).
However, the corresponding angular power spectrum is sourced by the geometric mean of linear and nonlinear matter power spectrum \citep{2012MNRAS.424.1647K}, reflecting the fact that lensing is sensitive to nonlinear structure growth at late times, while, presumably, intrinsic ellipticities have already built up in the early stages of galaxy formation, thus, in the linear regime. The corresponding reduced growth factor (normalized to unity today) is denoted by~$\bar D_+(\chi) \equiv D_+ (\chi) /a(\chi)$ \citep[e.g.][]{1998ApJ...508..483W,2003MNRAS.346..573L}.

\subsection{Biases to CMB lensing galaxy shape cross-correlation power spectra}

The power spectra of CMB lensing-galaxy shape cross-correlations given in equations~\eqref{eq:CMB_lensing_cosmic_shear_power_spectrum} and~\eqref{eq:CMB_lensing_itrinsic_alignment_power_spectrum} do not take into account that the CMB lensing potential is inaccessible to observations. When using real data only its estimate~\eqref{eq:CMB_lensing_potential_estimator} inferred from suitably weighted lensed CMB temperature maps is available. The lensed temperature field, however, contains all orders of the lensing potential gradient (and thus gradients of the Newtonian potential; cf. equation~\ref{eq:lensed_CMB_temperature}). Consequently, the CMB lensing galaxy shape cross-correlation acquires additional contributions beyond the two-point function. Focusing on the cosmic shear part (but similar expressions hold when the lensing induced ellipticity is replaced by the intrinsic galaxy ellipticity) this can be schematically written as
\begin{multline}
\left\langle \hat\phi \, \gamma \right\rangle 
 \sim \left\langle \tilde\Theta\tilde\Theta \, \gamma\right\rangle
 \sim 2\left\langle \Theta \nabla_a\Theta \right\rangle \left\langle \nabla^a \phi \, \gamma\right\rangle_\mathrm{A}
 + \left\langle \nabla_a\Theta \nabla_b \Theta \right\rangle \left\langle \nabla^a \phi  \nabla^b\phi  \, \gamma \right\rangle_\mathrm{B} \\
 + 2  \left\langle \Theta \nabla_a \nabla_b \Theta \right\rangle \left\langle \nabla^a \phi \nabla^b\phi \, \gamma \right\rangle_\mathrm{C}
 + \mathcal{O}\left[\left( \nabla^a\phi\right)^3 \, \gamma \right]
 \label{eq:schematic_expansion}
\end{multline}
where we have assumed that cosmic shear and unlensed CMB fluctuations are uncorrelated, i.e. we ignore the small correlation due to the integrated Sachs-Wolfe effect \citep{2016A&A...594A..21P}. Equation~\eqref{eq:schematic_expansion} incorporates consistently all terms of the perturbative expansion of the lensed CMB temperature (equation~\ref{eq:lensed_CMB_temperature}) up to second order in the lensing potential gradient.
From the harmonic space representation of term~A we recover the power spectrum~\eqref{eq:CMB_lensing_cosmic_shear_power_spectrum}. The other two terms constitute an additive bias. For any symmetric, especially a Gaussian, distribution the three-point functions are zero and these terms would vanish. However, nonlinear structure growth makes the distribution of matter fluctuations and likewise that of the gravitational potential skewed \citep[e.g.][]{2002PhR...367....1B}.

In order to compute the bias one has to express the correlators~B and~C  in harmonic space. There the three-point function is expressed by the bispectrum and the products of gradients become convolutions with a distinct wave vector geometry for each of the two terms:
\begin{multline}
 N^{\phi\gamma,\epsilon}_\mathrm{B}(\ell) = - A_\ell \int \frac{\dd^2 \ell_1}{(2\upi)^2} \int \frac{\dd^2 \ell_2}{(2\upi)^2} \,
 					g(\ellvec_1, \ellvec )
					\left[ \left( \ellvec_1 - \ellvec_2 \right) \cdot \ellvec_2 \right] \\
					\times
					\left[ \left( \ellvec_1 - \ellvec_2 \right) \cdot \left( \ellvec - \ellvec_2 \right) \right]
					C^{\Theta\Theta}_{\left|\ellvec_1 -\ellvec_2 \right|}
					B_{\phi\phi\gamma,\epsilon} \left( \ellvec_2, \ellvec - \ellvec_2, -\ellvec \right),
\label{eq:bias_B}
\end{multline}
\begin{multline}
 N^{\phi\gamma,\epsilon}_\mathrm{C}(\ell) = A_\ell \int \frac{\dd^2 \ell_1}{(2\upi)^2} \int \frac{\dd^2 \ell_2}{(2\upi)^2} \,
 					g(\ellvec_1, \ellvec )
					\left( \ellvec_1 \cdot \ellvec_2 \right) \\
					\times
					\left[ \ellvec_1\cdot \left( \ellvec - \ellvec_2 \right) \right] 
					C^{\Theta\Theta}_{\ell_1}
					B_{\phi\phi\gamma,\epsilon} \left( \ellvec_2, \ellvec - \ellvec_2, -\ellvec \right).
\label{eq:bias_C}
\end{multline}
Similar expressions have been derived by \citet{2016PhRvD..94d3519B} in the context of bispectrum induced biases to the reconstructed CMB lensing potential power spectrum.
Both terms are of similar magnitude but of different sign leading to a partial cancellation (cf. Figure~\ref{fig:bias_using_hyperextended_pt}). Therefore they cannot be treated separately and the actual bias to the power spectrum of CMB lensing galaxy shape cross-correlations is given by
\begin{equation}
 N^{\phi\gamma,\epsilon}_\ell =  N^{\phi\gamma,\epsilon}_\mathrm{B}(\ell) +  N^{\phi\gamma,\epsilon}_\mathrm{C}(\ell).
\end{equation}
For the (flat-sky) angular bispectrum~$B_{\phi\phi\gamma,\epsilon}$ we insert the Limber projection of the large-scale structure bispectrum \citep{2004MNRAS.348..897T}
\begin{multline}
 B_{\phi\phi\gamma,\epsilon} (\ellvec_1, \ellvec_2, \ellvec_3 ) =
	\int_0^\infty \dd\chi W_\mathrm{CMB}^2 (\chi) \, W_{\gamma,\epsilon} (\chi) \, \eta^3(\chi) \,\\
 	\times \ell_1^{-2}\ell_2^{-2}\, B_{\delta\delta\delta} ( \ellvec_1/\chi, \ellvec_2/\chi, \ellvec_3/\chi).
\end{multline}
Details on large-scale structure bispectrum computation are given in Appendix~\ref{sec:large_scale_structure_bispectrum}. In Figure~\ref{fig:equilateral_bispectra} we plot the equilateral angular bispectrum for both lensing induced and intrinsic ellipticities and note that the latter is surpassed in amplitude by more than one order of magnitude.

\section{Results}
\label{sec:results}

\begin{figure}
 \resizebox{\hsize}{!}{\includegraphics{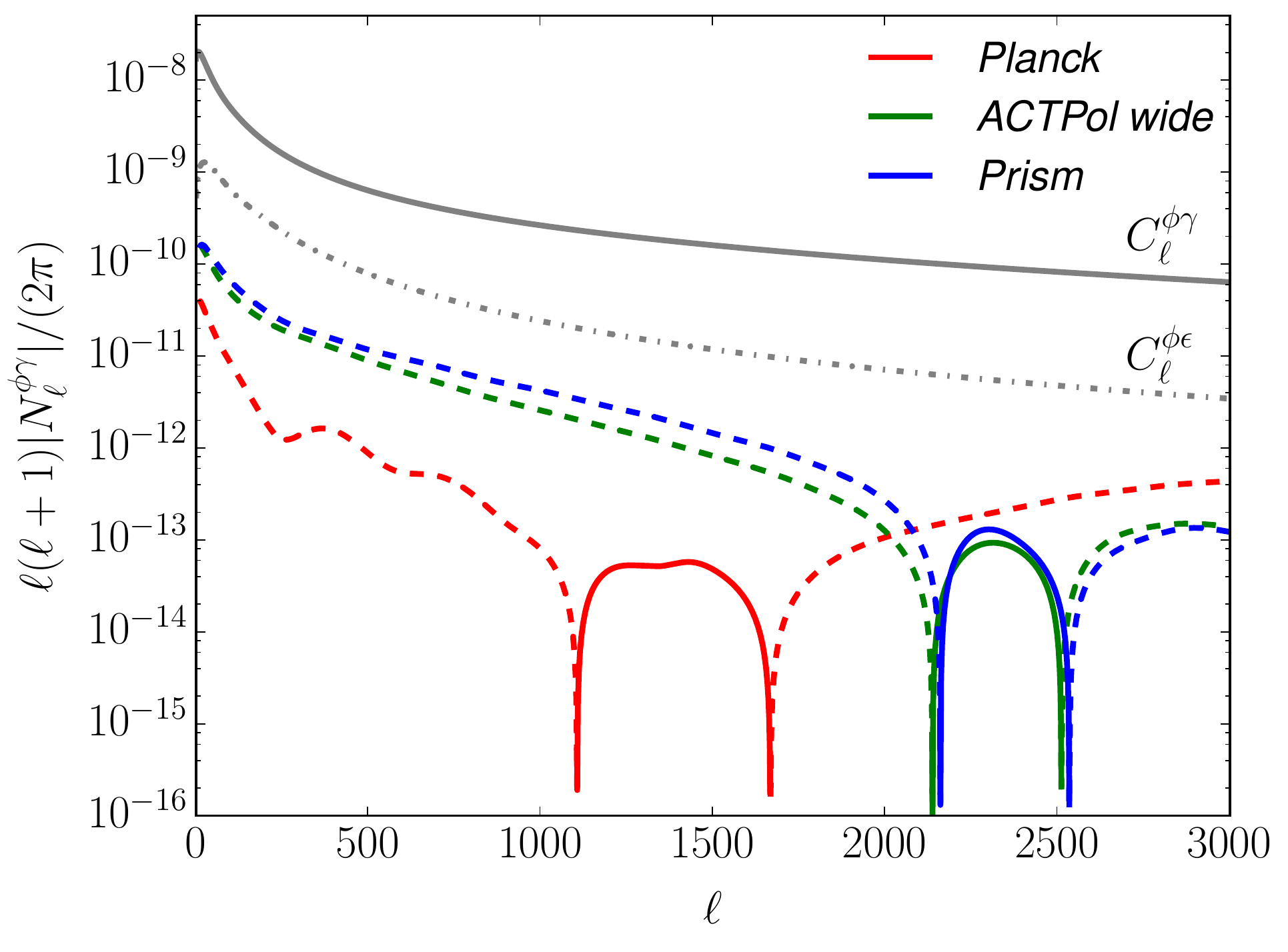}}
 \vspace{-15pt}
 \caption{CMB lensing-cosmic shear bias. Power spectra of CMB lensing-cosmic shear and CMB lensing-intrinsic galaxy ellipticity cross-correlations are shown as reference. Dashed lines indicate negative values.}
 \label{fig:CMB_lensing_cosmic_shear_bias}
\end{figure}

\begin{figure}
 \resizebox{\hsize}{!}{\includegraphics{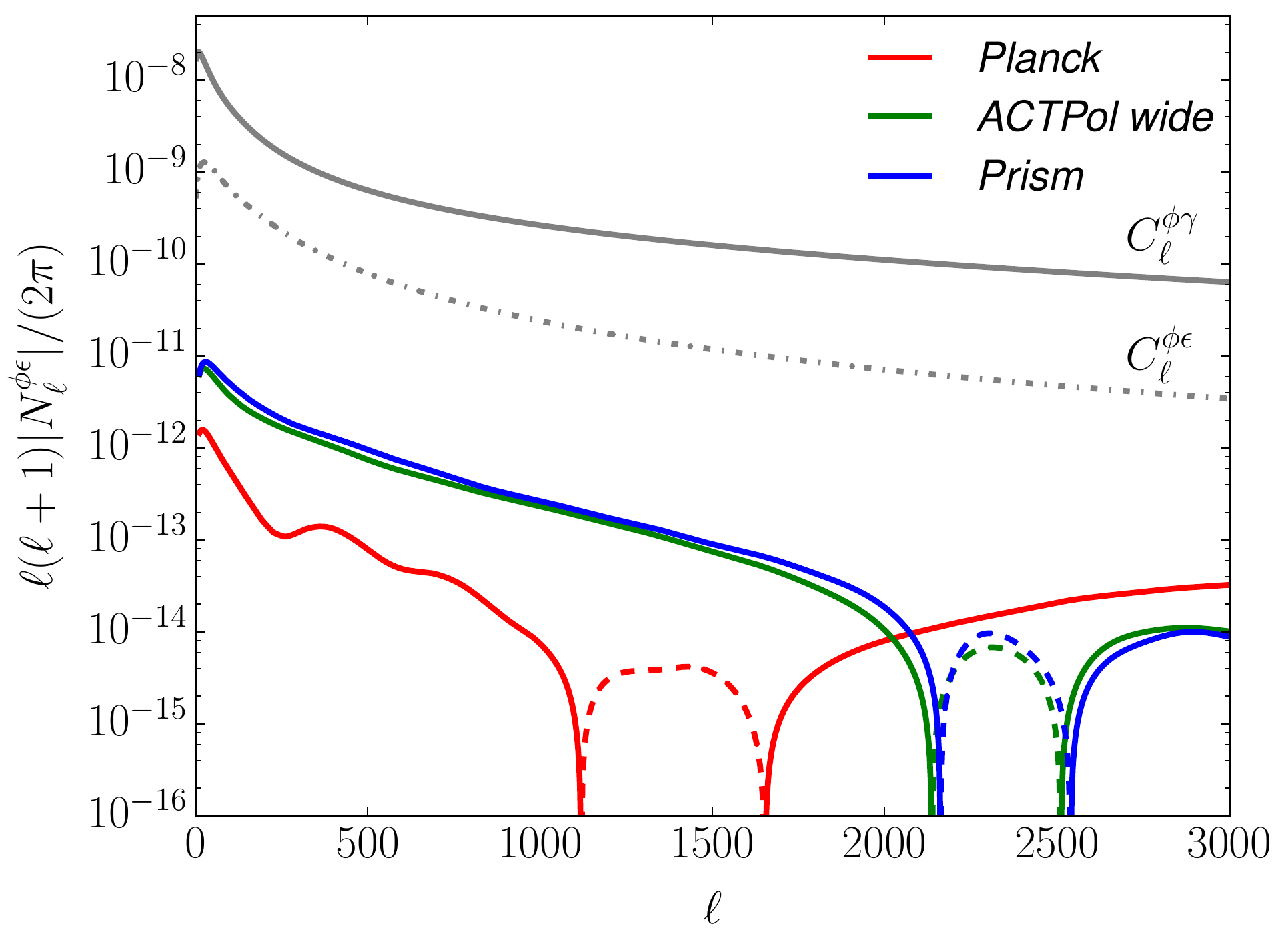}}
 \vspace{-15pt}
 \caption{CMB lensing-intrinsic galaxy ellipticity bias. Power spectra of CMB lensing-cosmic shear and CMB lensing-intrinsic galaxy ellipticity cross-correlations are shown as reference. Dashed lines indicate negative values.}
 \label{fig:CMB_lensing_ia_bias}
\end{figure}

We analyze the power spectrum bias for the three CMB experiments of Table~\ref{tab:CMBExperiments} in combination with galaxy shape measurements of a \textit{Euclid}-like cosmic shear survey. Figures~\ref{fig:CMB_lensing_cosmic_shear_bias} and~\ref{fig:CMB_lensing_ia_bias} show the power spectra of CMB lensing-cosmic shear and CMB lensing-intrinsic galaxy ellipticity cross-correlations, respectively, and the corresponding biases. Concentrating on Figure~\ref{fig:CMB_lensing_cosmic_shear_bias} first, we notice that the bias to the CMB lensing-cosmic shear cross-correlation is at least two orders of magnitude smaller than the power spectrum itself on all scales and irrespective of CMB data quality. Nevertheless, there are substantial differences in amplitude between the various CMB experiments. Broadly speaking the better the available CMB data the larger the bias on intermediate scales. For multipoles~$\ell \lesssim 2000$ the bias is about 100 times larger for \textit{ACTPol wide} and \textit{Prism} than for \textit{Planck}. On smaller angular scales, however, the situation is reversed and the bias obtained for \textit{Planck}-like data starts gently dominating; from multipoles~$\ell \sim 2500$ on the bias computed using highest CMB data quality is smallest. Generally the bias is negative for most angular scales but changes sign in a multipole band of width~$\Delta\ell\sim 500$. In case of \textit{Planck}-like data the change in sign occurs on larger angular scales (about 10\arcmin) than for the other two experiments where it almost coincides (at a scale of a roughly~5\arcmin). In fact the differences between the stage~III and stage~IV CMB experiments are much less pronounced compared to the results obtained for \textit{Planck}.

Moving on to Figure~\ref{fig:CMB_lensing_ia_bias} quite similar results are found. Though, there are important differences. First of all the bias is almost completely positive, i.e. its sign is reversed with respect to the CMB lensing-cosmic shear case. Furthermore its amplitude is roughly one order of magnitude smaller (cf. the proportion of the respective angular bispectra in Figure~\ref{fig:equilateral_bispectra}). Thus the bias merely amounts to one per~mil of the CMB lensing-cosmic shear power spectrum. Common to both biases is the scaling with improving CMB data and the characteristic zero-crossings.
\begin{figure}
 \resizebox{\hsize}{!}{\includegraphics{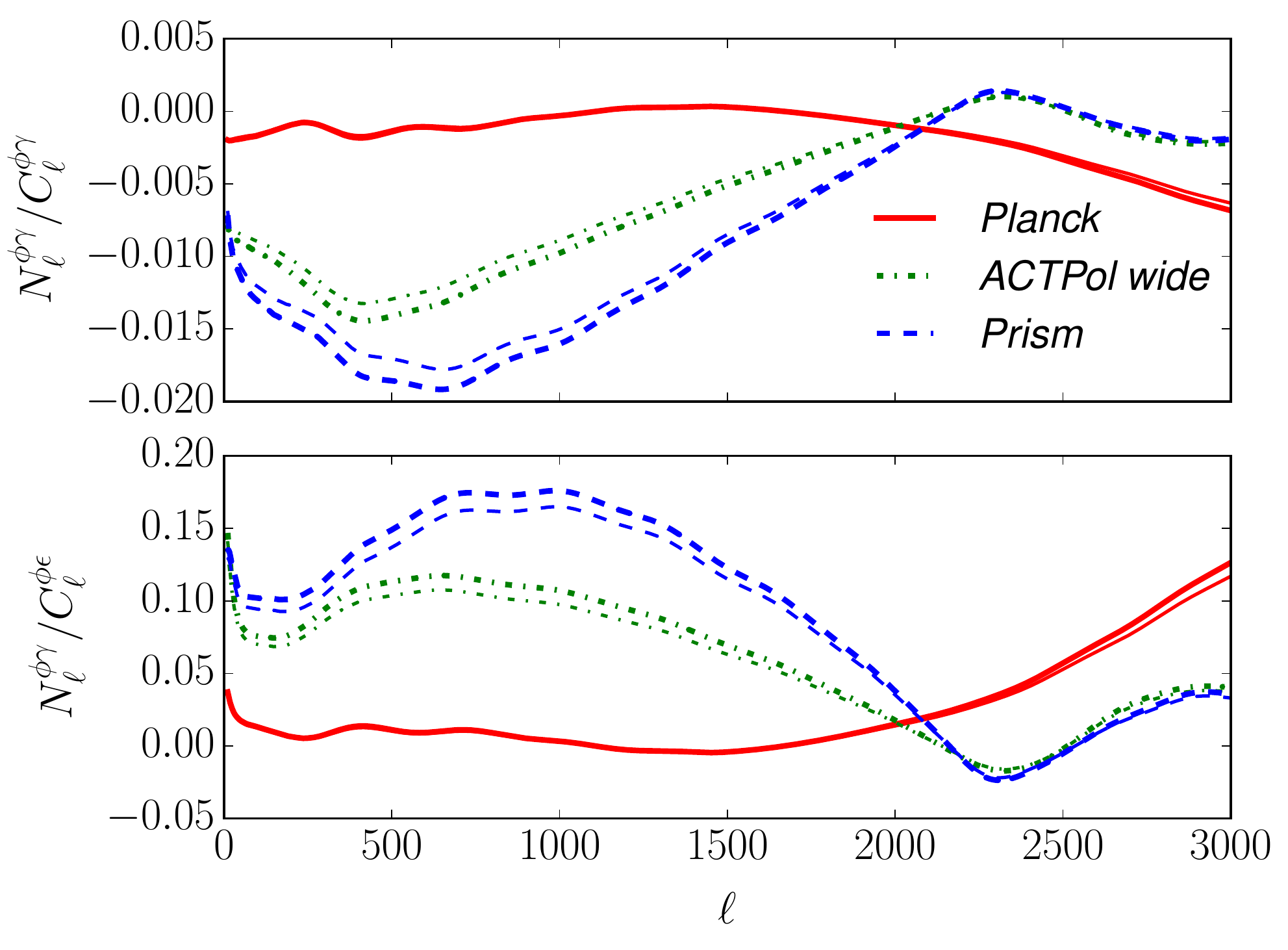}}
 \vspace{-15pt}
 \caption{Fractional contribution of the CMB lensing-cosmic shear bias to the power spectra of CMB lensing-cosmic shear cross-correlations (upper panel) and CMB lensing-intrinsic galaxy ellipticity cross-correlations (lower panel). Light curves indicate that the CMB lensing-intrinsic ellipticity bias is included.}
 \label{fig:ratio_bias_to_signal}
\end{figure}

For a more quantitative analysis we gather the ratios for the various combinations of biases and signals in Figure~\ref{fig:ratio_bias_to_signal}.
While the CMB lensing-cosmic shear power spectrum is hardly affected by the biases (cf. upper panel of Figure~\ref{fig:ratio_bias_to_signal}), they become important for correctly modeling CMB lensing-intrinsic galaxy ellipticity cross-correlations. The lower panel of Figure~\ref{fig:ratio_bias_to_signal} reveals that ignoring the CMB lensing-cosmic shear bias for high-quality CMB data may overestimate the CMB lensing-intrinsic alignment power spectrum  by up to 15~per~cent on intermediate scales; smaller angular scales are altered by less than 5~per~cent. Conversely, when using currently available \textit{Planck}-like data all but the smallest scales are unaffected; for multipoles~$\ell \gtrsim 2500$ the bias amounts to 5-10~per~cent. The effect of the bias is slightly mitigated by the fact that the  CMB lensing-intrinsic galaxy ellipticity bias has opposite sign (cf. Figures~\ref{fig:CMB_lensing_cosmic_shear_bias} and~\ref{fig:CMB_lensing_ia_bias}).
\begin{figure}
 \resizebox{\hsize}{!}{\includegraphics{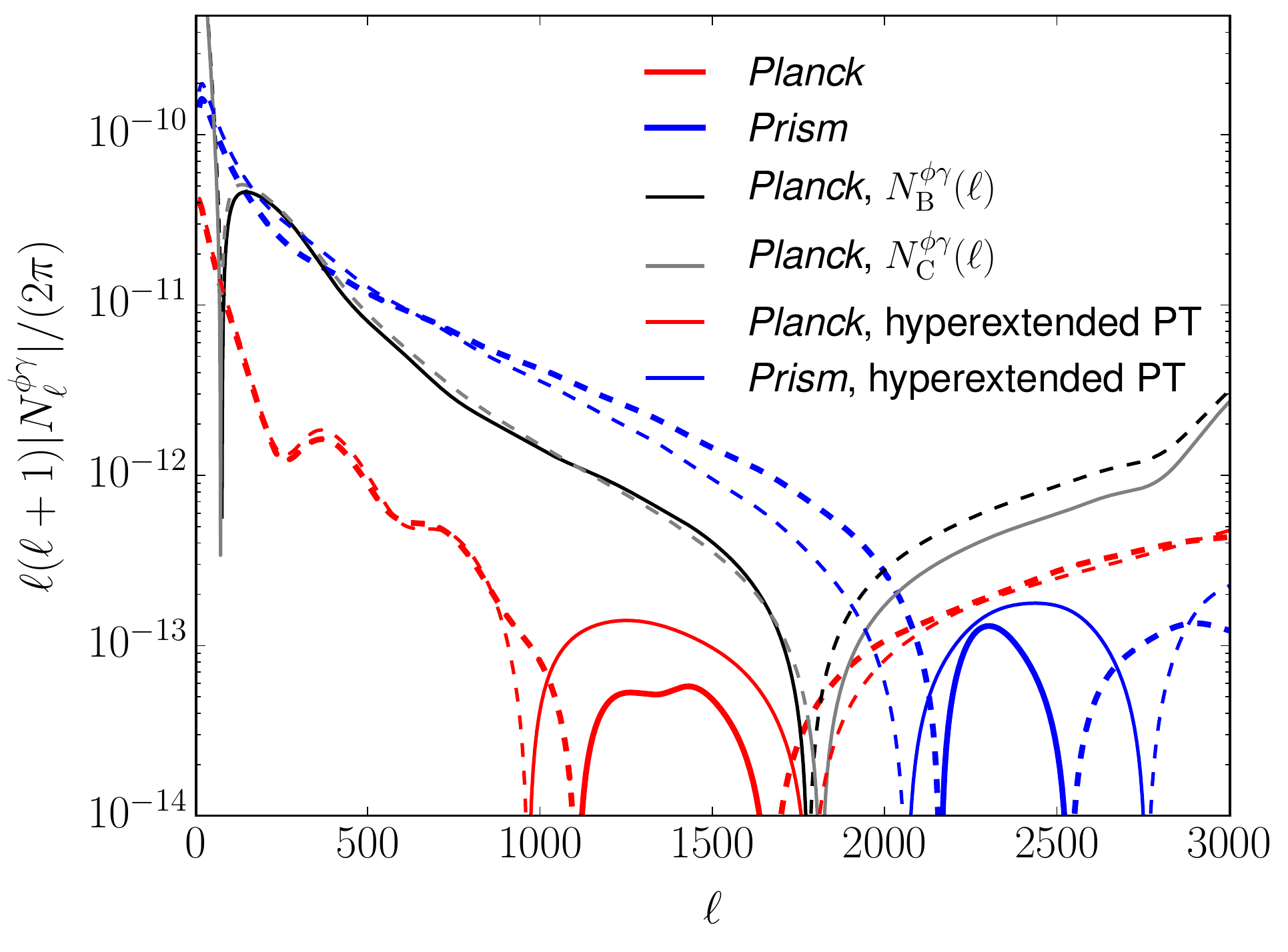}}
 \vspace{-15pt}
 \caption{CMB lensing-cosmic shear bias obtained using Eulerian (thick lines) and hyper extended (thin lines) perturbation theory for the bispectrum computation. As before negative values are indicated by dashed lines. The results for \textit{ACTPol wide}-like CMB data have been omitted for clarity. For the \textit{Planck} mission also the individual contributions to the bias are shown (thin grey lines).}
 \label{fig:bias_using_hyperextended_pt}
\end{figure}

For all results presented so far we have used Eulerian perturbation theory at tree-level to compute the large-scale structure bispectrum (see Appendix~\ref{sec:large_scale_structure_bispectrum}). We now investigate the impact of this choice on our findings. In our analysis we include contributions of small-scale matter fluctuations up to scales~$k_\mathrm{max} \sim 1 \, h \, \mathrm{Mpc}^{-1}$. The tree-level approximation, however, is only valid on scales larger than~$k \lesssim 0.1\, h\, \mathrm{Mpc}^{-1}$, whereas on smaller scales it underestimates the nonlinear power considerably (cf. Figure~\ref{fig:equilateral_bispectra}). These scales, however, can be accessed by either the halo model approach \citep{2002PhR...372....1C} or hyperextended perturbation theory \citep{1999ApJ...520...35S}. In Figure~\ref{fig:bias_using_hyperextended_pt} we compare the CMB lensing-cosmic shear bias obtained by using tree-level perturbation theory to that resulting from hyperextended perturbation theory. We concentrate on least and most precise CMB experiment, respectively; the results for a stage~III CMB survey do not differ. 
The main effect is to redistribute the power between the angular scales and to broaden the multipole band where the bias changes sign. However, the overall amplitude of the bias is not significantly enhanced by the increased small-scale power. This meets expectations because CMB lensing is most prominent on intermediate angular scales. Accordingly the filter function~\eqref{eq:estimator_weight_function} of the lensing potential estimator is particularly sensitive to these scales and it is exactly this filter function which is applied to the bispectrum when calculating the bias (cf. equations~\ref{eq:bias_B} and~\ref{eq:bias_C}). Thus given the general uncertainties inherent to models of nonlinear structure growth we believe that tree-level results are sufficiently accurate. This is in contrast to cosmic shear results, which are highly sensitive to small scales and mandatorily require advanced modeling of the large-scale structure bispectrum beyond tree-level \citep{2003MNRAS.344..857T,2014MNRAS.445.2918M}.

Finally, we investigate how the CMB lensing-galaxy shape cross-correlation biases, if uncorrected, affect measurements of the intrinsic alignment amplitude~$\mathcal{A}_\mathrm{I}$. Intrinsic ellipticity contributions are commonly determined from low-$z$ data and very little is known about alignments at~$z \gtrsim 1.2$ \citep[e.g.][and references therein]{2015MNRAS.453..682C}. Therefore CMB lensing-intrinsic ellipticity cross-correlations are expected to help constrain its amplitude at high redshift \citep{2014MNRAS.443L.119H,2015PhR...558....1T,2015MNRAS.449.2205K}. To this end correct modeling of the signal is mandatory. In order to quantify the systematical error~$\Delta\theta_\alpha$ which is induced by ignoring the biases in the observed power spectra we perturb the Fisher-approximated likelihood and arrive at 
\begin{equation}
 \Delta\theta_\alpha = F^{-1}_{\alpha\beta} \sum_\ell \frac{2\ell + 1}{2} C^{-1}_\ell \frac{\upartial C_\ell}{\upartial\theta_{\beta}}C^{-1}_\ell \Delta C_\ell
\end{equation}
\citep{2006MNRAS.366..101H,2008MNRAS.391..228A}.
The Fisher information matrix~$F_{\alpha\beta}$ \citep{1997ApJ...480...22T} of the set of cosmological parameters~$\theta_\alpha$ is constructed from the (uncorrected) power spectra~$C_\ell = C_\ell^{\phi\gamma} + C_\ell^{\phi\epsilon}$ and the bias contribution is given by the sum~$\Delta C_\ell = N^{\phi\gamma}_\ell + N^{\phi\epsilon}_\ell$. 
\begin{figure}
 \resizebox{\hsize}{!}{\includegraphics{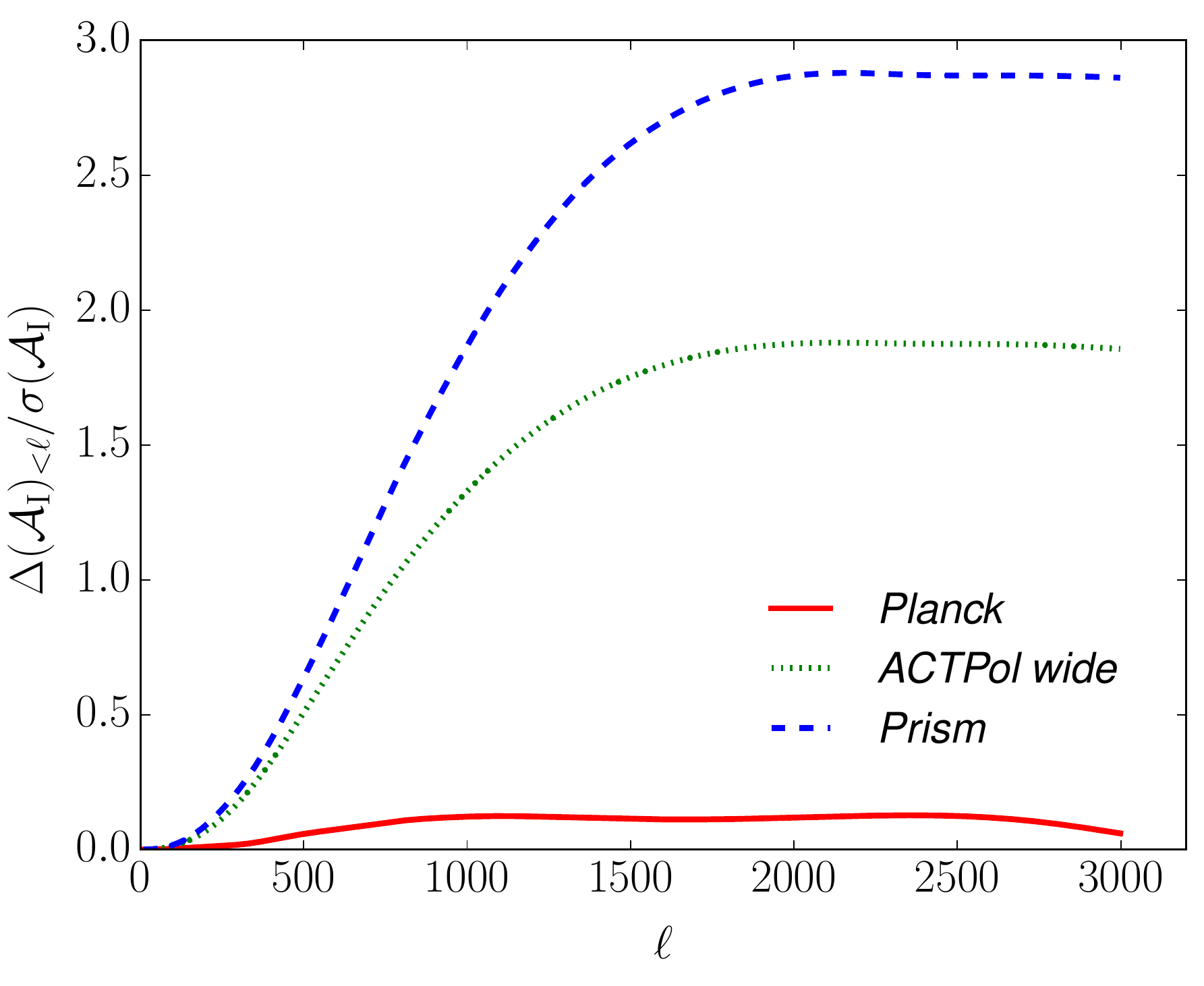}}
 \vspace{-15pt}
 \caption{Parameter estimation bias in the amplitude of intrinsic galaxy ellipticities~$\mathcal{A}_\mathrm{I}$, as fraction of the statistical error, caused by neglecting the CMB lensing-galaxy shape cross-correlation bias for multipoles smaller than~$\ell$.}
 \label{fig:parameter_estimation_bias}
\end{figure}
In Figure~\ref{fig:parameter_estimation_bias} we compare the estimation bias in the intrinsic alignment amplitude~$\Delta(\mathcal{A}_\mathrm{I})_{<\ell}$ to its statistical error~$\sigma(\mathcal{A}_\mathrm{I})=\sqrt{F^{-1}_{\mathcal{A}_\mathrm{I}\mathcal{A}_\mathrm{I}}}$. We assume that for multipoles larger than~$\ell$ the power spectrum biases are taken into account while for smaller multipoles they are entirely neglected. In addition to the intrinsic alignment amplitude the following parameters~$\left\lbrace \Omega_\mathrm{m}, \, \Omega_\mathrm{b}, \, \sigma_8, \, h, \, n_\mathrm{s}  \right\rbrace$ enter the Fisher matrix. We use prior information from six bin cosmic shear tomography of \textit{Euclid}-like lensing data and the respective CMB experiment as well. The prior treats both cosmological probes as statistically independent. With this set-up properties of intrinsic galaxy ellipticities can be determined at the per~cent level \citep{2017MNRAS.469.2760M}. Figure~\ref{fig:parameter_estimation_bias} shows that ignoring the power spectrum bias, which is almost exclusively negative (cf. Figures~\ref{fig:CMB_lensing_cosmic_shear_bias} and~\ref{fig:CMB_lensing_ia_bias}), overestimates the value for the alignment amplitude. For a \textit{Planck}-like experiment the bias is negligible, whereas it amounts to~$2\sigma$ and~$3\sigma$ for stage~III and stage~IV CMB experiments, respectively. Thus, for high-quality CMB data the power spectrum bias has necessarily to be accounted for in the data analysis.

\section{Conclusion}
\label{sec_conclusion}
We have computed the bias to the CMB lensing-galaxy shape cross-correlation, which is induced by the bispectrum of large-scale structures. Our analysis focuses on the lensing signal reconstructed from CMB temperature fluctuations, while its generalization to polarization based estimators is reserved for future work. There are two distinct contributions to the bias, which differ in magnitude and sign: the CMB lensing-cosmic shear bias is negative on almost all angular scales and about ten times larger than the CMB lensing-intrinsic galaxy ellipticity bias. Combining data from experiments that correspond to \textit{Euclid} and \textit{Planck} for galaxy and CMB observations, respectively, the bias is found to be negligible on all but the smallest scales~($\ell\gtrsim 2500$). When  \textit{Planck}-like observations are replaced by the increased precision of stage~III and stage~IV CMB experiments the bias is considerably enhanced and becomes appreciable on large and intermediate scales ($10\lesssim \ell \lesssim 2000$). Compared to the primary signal the bias is still small (about 1-2~per~cent) but it amounts to 10-15~per~cent of the CMB lensing-intrinsic alignment power spectrum. We have demonstrated that the additional power results in a systematically overestimated amplitude of intrinsic alignments at the~$2\sigma$ and~$3\sigma$-level in case of stage~III and stage~IV CMB observations, respectively. We therefore conclude that the power spectrum bias presented in this work is too small to help relax the tension between measured CMB lensing-cosmic shear cross-correlation power spectrum and its $\Lambda\mathrm{CDM}$ prediction but that it is compulsory to take the CMB lensing-galaxy shape cross-correlation bias into account when aiming at the properties of intrinsic alignments.


\bibliography{references}
\bibliographystyle{mnras}

\appendix
\section{Large-scale structure bispectrum}
\label{sec:large_scale_structure_bispectrum}

All observables presented in this work are projections of the Newtonian gravitation potential, which is related to the large-scale structure via the (comoving) Poisson equation
\begin{equation}
k^2 \Phi(\bmath k, a) =  \eta (a) \, \delta (\bmath k, a ), \qquad \eta(\chi) \equiv -\frac{3}{2} \frac{\Omega_\mathrm{m}H_0^2}{a(\chi)}.
\label{eq:Poisson_equation}
\end{equation}
Here we have directly stated its harmonic space version for convenience. The (equal-time) two- and three-point correlators of the density contrast define its power spectrum and bispectrum, respectively
\begin{gather}
 \left\langle\delta(\bmath k_1)\, \delta^*(\bmath k_2) \right\rangle 
 							= (2\upi)^3 \, \delta_\mathrm{D}(\bmath k_1 - \bmath k_2) \,P_{\delta\delta} (k_1),\\
 \left\langle\delta(\bmath k_1)\, \delta(\bmath k_2)\, \delta(\bmath k_3) \right\rangle 
 				= (2\upi)^3 \, \delta_\mathrm{D} ( \bmath k_1 + \bmath k_2 + \bmath k_3 )\, 
							B_{\delta\delta\delta} (\bmath k_1, \bmath k_2, \bmath k_3).
\end{gather}
The impact of nonlinear clustering on the power spectrum is accounted for by using the revised halo model approach of \citet{2012ApJ...761..152T} for the additional power on small scales. To distinguish between linear and nonlinear matter power spectra we use~$P^\mathrm{lin}_{\delta\delta}$ and~$P^\mathrm{nl}_{\delta\delta}$, respectively. Nonlinear structure growth does not only enhance the small-scale power but also skews the distribution function of matter fluctuations, which we assume to be initially Gaussian. Thus, we do not consider primordial non-Gaussianities \citep[see][for reviews]{2010CQGra..27l4010K,2010AdAst2010E..71Y} as currently favoured values of its amplitude \citep{2016A&A...594A..17P} suggest that its contributions to the matter bispectrum are negligible.

For the computation of the bispectrum we employ Eulerian perturbation theory truncated at tree-level \citep{2002PhR...367....1B,2011PhRvD..83h3518M}
\begin{equation}
 B_{\delta\delta\delta} (\bmath k_1, \bmath k_2, \bmath k_3 ) = \sum_{\substack{i,j =1,2,3\\ i \neq j}} F ( \bmath k_i, \bmath k_j ) \,
 		P^\mathrm{nl}_{\delta\delta}(k_i) \, P^\mathrm{nl}_{\delta\delta} (k_j )
 \label{eq:lss_bispectrum}
\end{equation}
where the mode-coupling kernel is given by
\begin{equation}
 F(\bmath k_1, \bmath k_2)= \frac{5}{7} + \frac{1}{2}\left( \frac{k_1}{k_2} + \frac{k_2}{k_1}\right) \mu + \frac{2}{7} \, \mu^2,
 		\quad \mu \equiv \cos \sphericalangle \left(\bmath k_1, \bmath k_2\right).
\end{equation}
We insert the nonlinear power spectrum in equation~\eqref{eq:lss_bispectrum} in order to extend the applicability of the tree-level expressions to smaller scales \citep{2001MNRAS.325.1312S}. Beyond scales of~$k \gtrsim 0.1 \, h \, \mathrm{Mpc}^{-1}$ more elaborated methods have to be invoked as discussed in Section~\ref{sec:results}.
\begin{figure}
 \resizebox{\hsize}{!}{\includegraphics{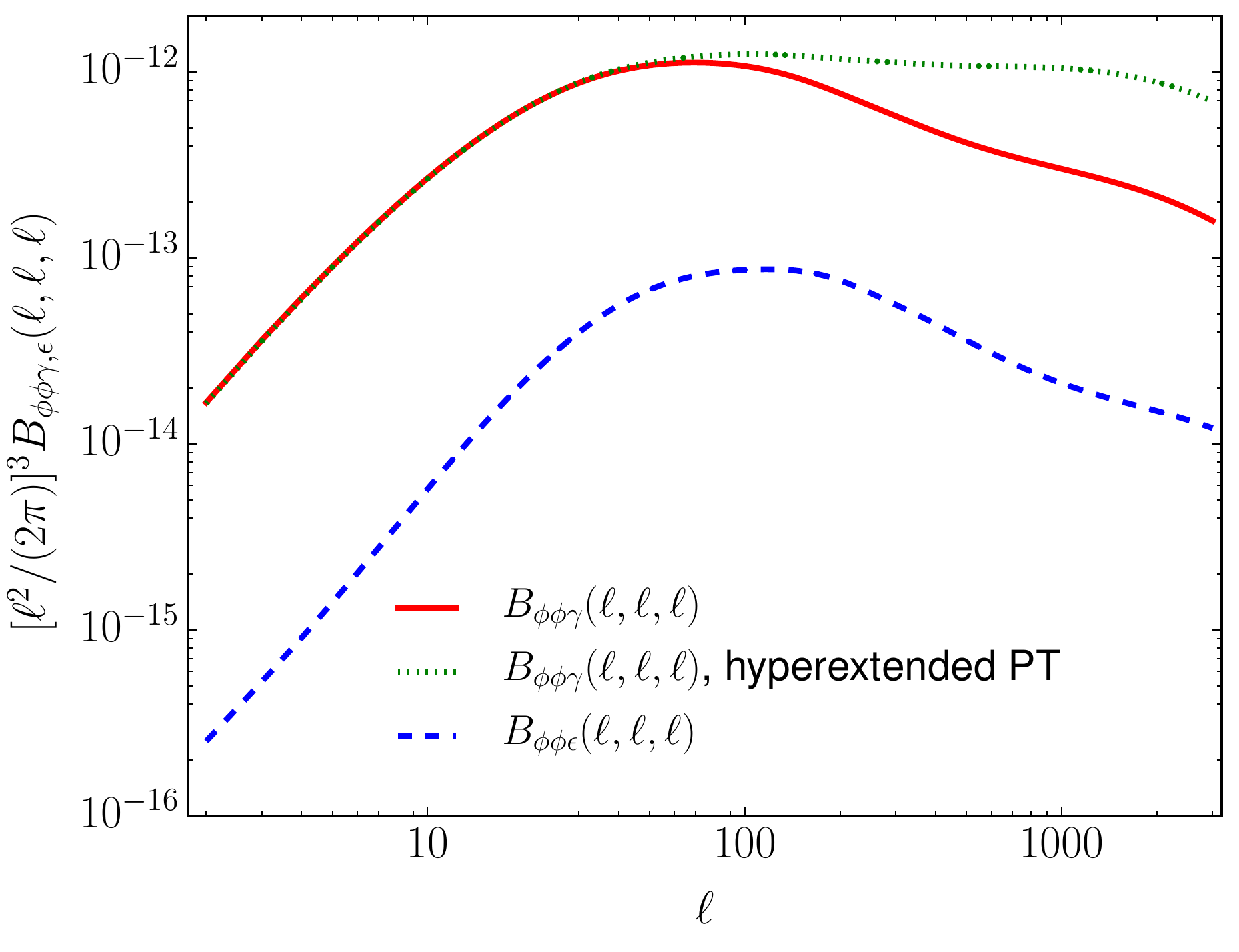}}
 \vspace{-15pt}
 \caption{Equilateral angular bispectra of CMB lensing-galaxy shape cross-correlation. For the CMB lensing-cosmic shear cross-correlation also the bispectrum obtained using hyperextended perturbation theory is shown.}
 \label{fig:equilateral_bispectra}
\end{figure}
The resulting differences in the angular projected bispectra can be seen in Figure~\ref{fig:equilateral_bispectra}.

\bsp

\label{lastpage}

\end{document}